\documentclass[aps,prl,reprint,showpacs,floatfix,superscriptaddress]{revtex4-1}
\usepackage[english]{babel}
\usepackage[utf8]{inputenc}
\usepackage{physics}
\usepackage{graphicx}
\usepackage{fancyhdr}
\usepackage[legalpaper, margin=1in, includefoot]{geometry}

\begin{document}

\title{Laser Field Initiation of Higher Order Poles of S-Matrix-Optical Realization of Field Theoretic Models*}

\author{G.S. Agarwal$^{1,2}$ \\
        \small $^{1}$Physical Research Laboratory, Navrangpura, Ahmedabad, India \\
        \small $^{2}$Jawaharlal Nehru Center for Advanced Scientific Research, Bangalore, India \\
}

\begin{abstract} 
\noindent We discuss the possibility of converting a simple pole in the radiative decay
of a state into a pole of higher order by using resonant electromagnetic fields.
This process of creation of higher order pole is controllable by the intensity
of the laser field. We use density matrix and Liouville space and present the
modification of the Lorentzian line shapes (Breit-Wigner formula) for example
to ones involving square of Lorentzian and derivatives of Lorentzians. \end{abstract}

\maketitle

\section{INTRODUCTION}
\thispagestyle{fancy}
\fancyfoot{}
\fancyfoot[L]{{\footnotesize ---------------------------------------------------------------------------------------------------------------------------------------------------------------------*Published in  “Frontiers of Quantum Optics and Laser Physics”, p.155-165, ed. S.Y. Zhu, M.S. Zubairy and M.O. Scully (Springer, 1997). This work on higher order poles of S matrix has close connection to the exceptional point physics.  Thus this work brings out how the exceptional point physics in active systems can be manipulated by laser field. See also G. S. Agarwal, Quantum Optics, Cambridge University Press, 2012, Section 17.3.1.}}

In a classic paper Goldberger and Watson [1] considered the possibility that the decay
law for an unstable particle can be more complex than a simple exponential. They showed
the possibility of the existence of the poles of S-matrix which were not necessarily simple
poles. Since then, higher order poles have been extensively studied. Recently, there is revival
[2,3] of interest in such studies and in particular Bhamathi and Sudarshan have analyzed
several field theoretic models like Friedrich-Lee model, cascade model and their extensions.
They examine the spectrum (complex) of eigenvalues for such models. A related question is
how the Breit-Wigner line shape formula is modified if S-matrix possess higher order poles.

In this paper we examine the possibility of creation of the higher order poles using
laser fields. We consider the decay of say excited state of an atom. Normally this decay
is described by the Wigner-Weisskopf theory which leads to exponential decay law.
We next discuss the case when the excited state is coupled to another state by a resonant
electromagnetic field. In such a case we show that for appropriate value of the intensity of
the laser field the corresponding spectral function has a pole of order two. We calculate the
resulting line shape and discuss the line narrowing etc. We emphasize that we work within
the framework of density matrices and hence we work in Liouville space rather than in
Hilbert space. We present optical realization of various field theoretic models.

Consider the decay of the state $\ket{1}$ into the states $\ket{3}$ and $\ket{2}$ at the rates 2$\gamma_1$ and 2$\gamma_2$
respectively as shown in Fig.1 (with $G_l=0, \triangle_l = 0$). It is well known that the rate of decay
of the population in $\ket{1}$ is given by
\begin{equation}
    \rho_{11}(t)=\rho_{11}(0)exp(-2(\gamma_1+\gamma_2)t).
\end{equation}
Here $\rho$ is the density matrix of the atom. The spectrum of the spontaneously emitted photons
will consist of two Lorentzians centered at $\omega_{13}$ and $\omega_{12}$ with a half width $(\gamma_1+\gamma_2)$.
Let us concentrate on the emission on the transition $\ket{1} \leftrightarrow \ket{3}$. The spectrum will be described by
the well-known form
\begin{equation}
    S(\omega)=\frac{\gamma_{1}/\pi}{(\gamma_{1}+\gamma_{2})^{2}+(\omega-\omega_{13})^{2}}.
\end{equation}
Note that $\gamma_{2}$ will be zero if the decay channel $\ket{1} \rightarrow \ket{2}$ is not allowed. We will discuss how
the laser fields could be used to modify significantly the results predicted by (1) and (2).

\section{LIOUVILLE SPACE FORMULATION OF DECAY}

We next recall how the spectrum is calculated in the density matrix framework [4]. We
have included this material for completeness so that our discussion in subsequent sections
can be followed by the non-Quantum optics practitioners. Consider a system with two states
$\ket{1}$ and $\ket{3}$ interacting with the vacuum of the electromagnetic field. The Hamiltonian can be written in the form
\begin{align}
&H=\hbar\omega_{13}\ket{1}\bra{1}+\sum_{ks}\hbar\omega_{ks}a_{ks}^{\dagger}a_{ks}+V_{13} \notag \\
&V_{13}=\sum_{ks}(\hbar g_{ks}a_{ks}^{\dagger}\ket{1}\bra{3}+h.c.).
\end{align}
The vacuum modes are characterized by the propagation index $\overrightarrow{k}$ and the polarization index $s$. 
The $a_{ks}, a_{ks}^{\dagger}$ represent annihilation and creation operators for the mode $\overrightarrow{k}s$. The $V_{13}$
describes the decay of $\ket{1}$ to $\ket{3}$. The $g_{ks}$ is the coupling constant between the field mode
and the atom. We use the weak coupling assumption and the flat nature of the density of
states of the electromagnetic vacuum to eliminate the degrees of freedom associated with
the field vacuum. We derive an equation for the density matrix of the atomic system alone
which can be written in the form
\begin{equation}
\frac{\partial\rho}{\partial t}=L\rho
\end{equation}
or in terms of the components as
\begin{align}
\dot{\rho}_{11}&=-2\gamma_{1}\rho_{11}, \notag \\
\dot{\rho}_{13}&=-i\omega_{13}\rho_{13}-\gamma_{1}\rho_{13}, \notag \\
\dot{\rho}_{33}&=2\gamma_{1}\rho_{11},\quad etc.,\:2\gamma_{1}=\sum_{ks}|g_{ks}|^{2}\delta(\omega_{13}-\omega_{ks}).
\end{align}
This yields steady state as well as transient behavior.
The spectrum of radiation is
related to the Fourier transform of the two time dipole correlation function, for example
in the above case to
\begin{equation}
S(\omega)=\frac{1}{\pi}Re[S(z)|_{z=+i\omega}],
\end{equation}
\begin{align}
S(z)&\equiv\int_{0}^{\infty}d\tau e^{-z\tau}\left\langle A_{13}(t+\tau)A_{31}(t)\right\rangle , \notag \\
A_{13}&=A_{31}^{\dagger}=\ket{1}\bra{3}.
\end{align}
The poles of $S(z)$ determine the spectrum. For the standard problem $S(z)$ has simple poles.

The two time correlation function is calculated from the solution of (4) and by using
the quantum regression theorem. For completeness, we state what it means. We write the
solution of (4) as
\begin{equation}
\rho_{\alpha\beta}(t+\tau)=\sum_{m,n}G_{\alpha\beta,mn}(\tau)\rho_{mn}(t).
\end{equation}
It should be borne in mind that in the Liouville space $\rho_{\alpha\beta}$ is an element of
the column matrix. We can rewrite (8) as
\begin{equation}
\left\langle A_{\beta\alpha}(t+\tau)\right\rangle =\sum_{m,n}G_{\alpha\beta,mn}(\tau)\left\langle A_{nm}(t)\right\rangle ,
\end{equation}
then the quantum regression theorem leads to two time correlation function:
\begin{align}
\left\langle A_{\beta\alpha}(t+\tau)A_{pq}(t)\right\rangle &\equiv\sum_{m,n}G_{\alpha\beta,mn}(\tau)\left\langle A_{nm}(t)A_{pq}(t)\right\rangle  \notag \\
&=\sum_{m,n}G_{\alpha\beta,mn}(\tau)\left\langle A_{nq}(t)\right\rangle \delta_{mp} \notag \\
&=\sum_{m,n}G_{\alpha\beta,mn}(\tau)\delta_{mp}\rho_{qn}(t).
\end{align}
On using (10) in (6) it is clear that $S(z)$ is related to the Laplace transform of $G(\tau)$ or to
$(z-L)^{-1}$. Generally, the Liouvilliean matrix relevant for the calculation of (10) decomposes
in block diagonal form and only a part of $L$ determines the decay or the spectral line shapes.
For the two level example, the correlation function is essentially determined by a single equation for $\rho_{13}$. If there is more than one decay channel, then additional terms appear in
(5), for example, for the case shown in Fig.1, $\gamma_1$ should be replaced by $(\gamma_1+\gamma_2)$ in the
two first equations in (5).

\begin{figure}[h]
\includegraphics[width=7cm]{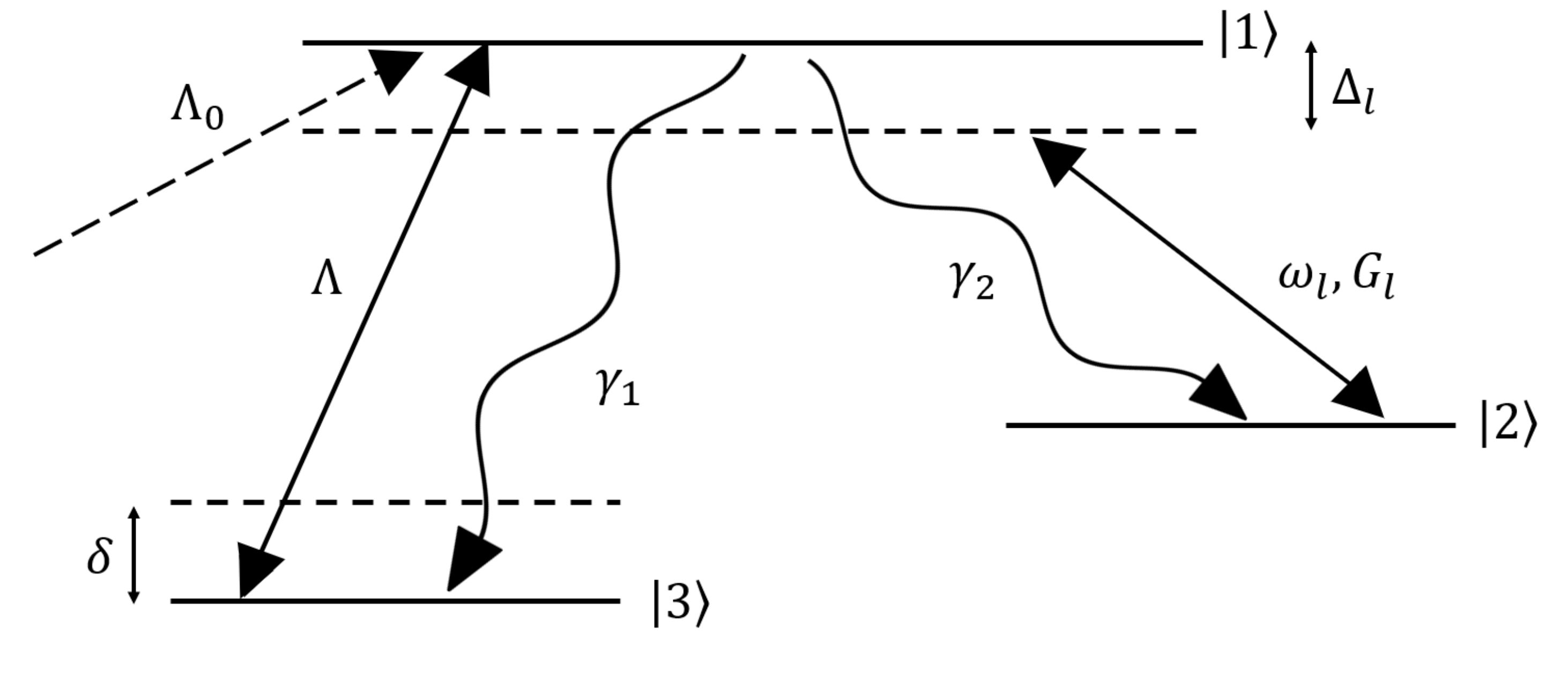}
\caption{Schematic illustration of the scheme that leads to the creation of poles of order two in
the decay of the state $\ket{1}$; which could be pumped in two different ways either from the state $\ket{3}$
or from a state outside the system. This provides the realization of the extended Friedrich-Lee model.}
\centering
\end{figure}

\section{CREATION OF A DOUBLE POLE}

We next demonstrate how by using external electromagnetic fields we can convert simple
poles of $L$ into poles of higher order. For this purpose, we consider the application of an
electromagnetic field that is tuned close to the transition frequency $\omega_{12}$ [Fig. 1.
$\Lambda_0=0, \Lambda\neq0, G_l\neq0]$. The Hamiltonian describing this system can be written as
\begin{equation}
H=\hbar\omega_{13}\ket{1}\bra{1}+\hbar(\omega_{13}-\omega_{12})\ket{2}\bra{2}+H_{ext}+V_{12}+V_{13},
\end{equation}
where $V_{\alpha \beta}$ describes the decay on the transition $\ket{\alpha}\rightarrow\ket{\beta}$ and where
\begin{equation}
H_{ext}=-\hbar(G_{l}e^{-i\omega_{l}t}\ket{1}\bra{2}+h.c.),
\end{equation}
\begin{equation}
G_{l}=(\overrightarrow{d}_{12}\cdot\overrightarrow{E}_{l}/\hbar).
\end{equation}
The parameter $2G_l$ is the Rabi frequency of the field and is a measure of the strength of
the laser field applied on the transition $\ket{1}\leftrightarrow\ket{2}$.
The Hamiltonian (11) is time-dependent. However one can make a canonical transformation to reduce it to a time-independent
Hamiltonian. In the special case $V_{12}\rightarrow0$ the model (11) is equivalent to the extended
Friedrich-Lee model. We have thus produced a realization of a field-theoretic model in the
context of atoms interacting with laser fields. In our case lasers are used to control the
decay process. Note that we have two control parameters $\omega_l$ and $G_l$, to manipulate the nature of the poles of $L$. The situation shown in Fig. 1 is realizable in many atoms, molecules
dopants in solid matrices, etc. For example, in $^{87}Rb$ vapor, the states $\ket{1}$, $\ket{2}$ and $\ket{3}$ could be
the states $^{5}P_{\frac{3}{2}},{}^{5}S_{\frac{1}{2}},F=2$ and $^{5}S_{\frac{1}{2}},F=1$, respectively. We eliminate the optical frequencies
by making canonical transformations $\rho_{13}\rightarrow\rho_{13}e^{-i\omega_{13}t},\rho_{12}\rightarrow\rho_{12}e^{-i\omega_{l}t}$ etc. After canonical
transformations and after eliminating vacuum degrees of freedom using the master equation
techniques the density matrix equations read [5]
\begin{align}
&\dot{\rho}_{11}=-2(\gamma_{1}+\gamma_{2}+\Lambda)\rho_{11}+2\Lambda\rho_{33}+iG_{l}\rho_{21}-iG_{l}^{*}\rho_{12},\notag \\
&\dot{\rho}_{22}=2\gamma_{2}\rho_{11}-iG_{l}\rho_{21}+iG_{l}^{*}\rho_{12},\notag \\
&\dot{\rho}_{21}=-(\Gamma_{21}-i\Delta_{l})\rho_{21}-iG_{l}^{*}\rho_{22}+iG_{l}^{*}\rho_{11},\notag \\
&\dot{\rho}_{31}=-\Gamma_{31}\rho_{31}-iG_{l}^{*}\rho_{32},\notag \\
&\dot{\rho}_{32}=-(\Gamma_{32}+i\Delta_{l})\rho_{32}-iG_{l}\rho_{31}.
\end{align}
Here we have also included a pumping parameter $\lambda$ to pump the population from the level
$\ket{3}$ to $\ket{1}$. The $\Gamma_{\alpha\beta}^{'s}$ give the decay of off-diagonal elements $\rho_{\alpha\beta}^{'s}$ of the density matrix and
are given by
\begin{align}
&\Gamma_{31}=\gamma_{1}+\gamma_{2}+2\Lambda,\Gamma_{32}=\Lambda,\notag \\
&\Gamma_{21}=\gamma_{1}+\gamma_{2}+\Lambda,\Delta_{2}=\omega_{12}-\omega_{l}.
\end{align}
From (14) and the quantum regression theorem we derive coupled equations for two time atomic correlation functions
\begingroup\makeatletter\def\f@size{8}\check@mathfonts
\def\maketag@@@#1{\hbox{\m@th\large\normalfont#1}}%
\begin{equation}
\left\{ \frac{d}{d\tau}+\left(\begin{array}{cc}
\Gamma_{31} & iG_{l}^{*}\\
iG_{l} & \Gamma_{32}+i\Delta_{2}
\end{array}\right)\right\} \left(\begin{array}{c}
\left\langle A_{13}(t+\tau)A_{31}(t)\right\rangle \\
\left\langle A_{23}(t+\tau)A_{31}(t)\right\rangle 
\end{array}\right)=0.
\end{equation} \endgroup
These are to be solved subject to initial conditions
\begin{equation}
\left\langle A_{13}A_{31}\right\rangle =\rho_{11},\left\langle A_{23}A_{31}\right\rangle =\rho_{12},
\end{equation}
which in turn are determined from the steady state solution of (14). Clearly the poles of $L$
that determine the spectral characteristics are given by
\begin{equation}
P(z)=(z+\Gamma_{31})(z+\Gamma_{32}+i\Delta_{l})+|G_{l}|^{2}.
\end{equation}
The zeroes of (18) for $\Delta_{l}=0$ are shown in Fig. 2. The conditions under which $P(z)$ has
double zero are
\begin{equation}
\Delta_{l}=0,(\Gamma_{32}-\Gamma_{31})^{2}=4|G_{l}|^{2}.
\end{equation}
The double zero $z_0$ occurs at the bifurcation point in Fig. 2
\begin{equation}
z_{0}=-\frac{1}{2}(\Gamma_{31}+\Gamma_{32}).
\end{equation}
We therefore conclude [6] that a simple pole can be converted into a double pole in a laboratory
experiment by applying an electromagnetic field resonant with the transition $\ket{1}\leftrightarrow\ket{2}$ and
with Rabi frequency equal to $|\Gamma_{31}-\Gamma_{32}|$.

\begin{figure}[h]
\includegraphics[width=7cm]{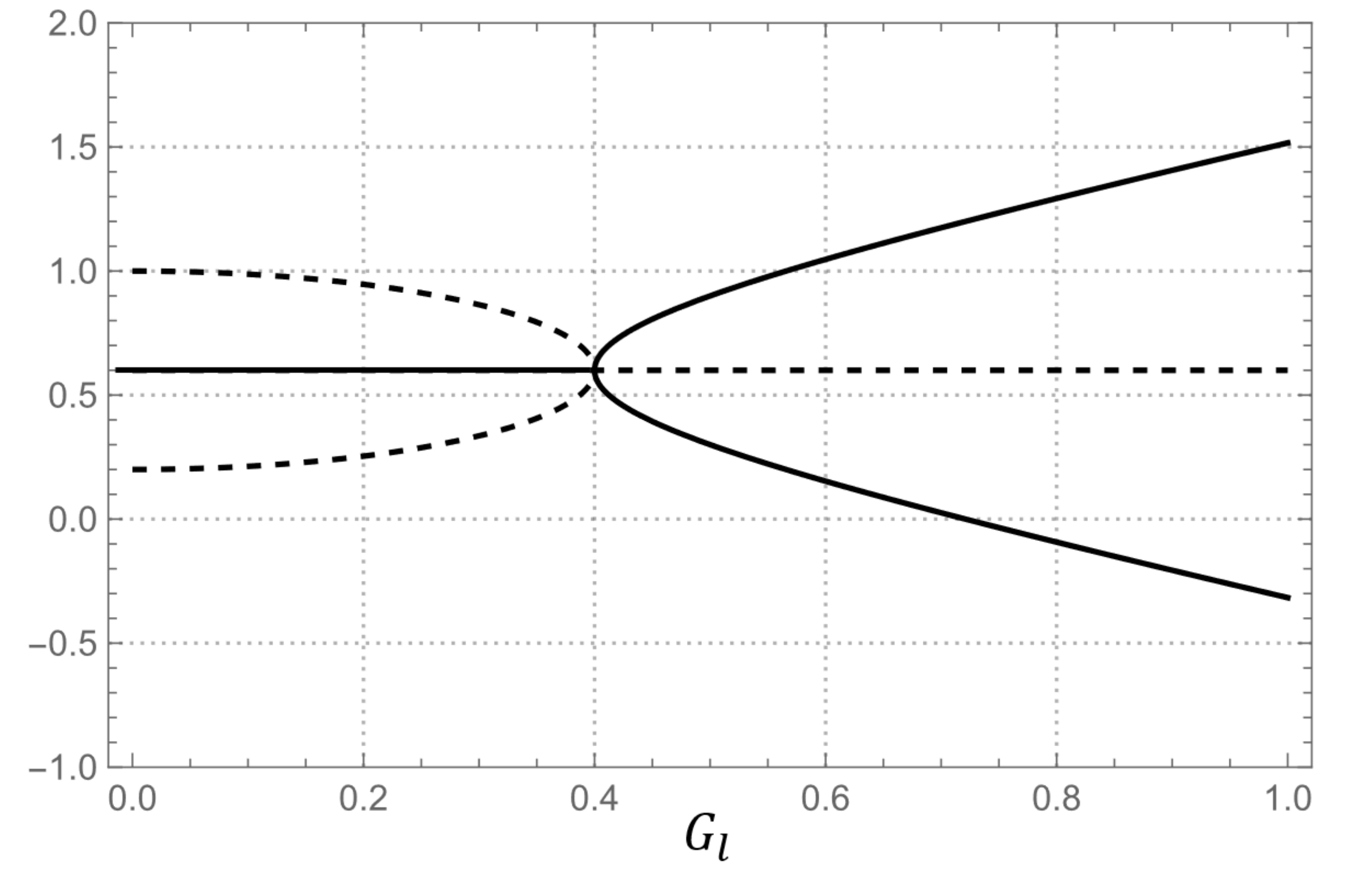}
\caption{Motion of the zeroes of (18) for $\Gamma_{31}=1,\Gamma_{32}=0.2$. Note the presence of the bifurcation
point. This is precisely the point where we create a pole of order two. The solid curve represents
$Im(z)+0.6$ whereas the dashed curve gives $Re(z)$.}
\centering
\end{figure}

\section{LINE SHAPES AND DOUBLE POLES}

The line shape can be calculated from the solution of (16) and (6):
\begin{equation}
S(\omega)\equiv\rho_{11}Re[\frac{(\gamma_{2}+\Gamma_{32}-i\delta)}{(\Gamma_{31}-i\delta)(\Gamma_{32}-i\delta)+|G_{l}|^{2}}]
\end{equation}
which under the double pole condition $2|G_l|=|\Gamma_{31}-\Gamma_{32}|$ reduces to
\begin{align}
S(\omega)&=\rho_{11}Re[\frac{\gamma_{2}+\Gamma_{32}-i\delta}{(-i\delta+\gamma_{0})^{2}}] \notag \\
&=\rho_{11}\frac{\delta^{2}(\gamma_{1}+2\Lambda)+\gamma_{0}^{2}(\gamma_{2}+\Lambda)}{(\delta^{2}+\gamma_{0}^{2})^{2}},\notag \\
\gamma_{0}&=\frac{1}{2}(\gamma_{1}+\gamma_{2}+3\Lambda).
\end{align}
This is the modification of the line shape formula. Note the double hump structure of the
line shape. Note further the sensitiveness of $S(\omega)$ to the pumping parameter $\Lambda$. In the limit $\gamma_{2}\rightarrow0$ and $\Lambda\ll\gamma_{1}$, (22) reduces to
\begin{equation}
S(\omega)\equiv\rho_{11}\frac{\gamma_{1}(\delta^{2}+\frac{\gamma_{1}}{4}\Lambda)}{(\delta^{2}+\frac{\gamma_{1}^{2}}{4})^{2}}
\end{equation}
It is also interesting to note, that the scale parameter is now $(\gamma_{1}/2)$ rather than $\gamma_{1}$.
Thus the total line shape is a sum of (a) Square of the Lorentzian (b) derivative of the Lorentzian $(\zeta/(\zeta+\gamma_{0})^{2}\equiv-\zeta\frac{\partial}{\partial\zeta}(\frac{1}{\zeta+\gamma}))$.

It is possible to consider an alternate model of pumping obtained by setting $\Lambda=0$ in Fig. 1.
Assuming that $\gamma_2=0$, one can show that instead of (23) the spectral line shape is now given by
\begin{equation}
S(\omega)\equiv\frac{\gamma_{1}\rho_{11}\delta^{2}}{(\delta^{2}+\frac{\gamma_{1}^{2}}{4})^{2}}=(-\delta\frac{\partial}{\partial\delta})\frac{(\gamma_{1}/2)\rho_{11}}{(\delta^{2}+\gamma_{1}^{2}/4)}
\end{equation}
which is shown in Fig. 3.
The figure also shows for comparison the Breit-Wigner formula (2)
Note the double hump structure of the line shape. The maxima now occur at $\delta=\pm\gamma_{1}/2$.
From Eq. (14) we can also compute the time dependence of $\rho_{11}(t)$ under the condition of a double pole. The result is
\begin{equation}
\rho_{11}(t)=(1-\frac{\gamma_{1}t}{2})^{2}e^{-\gamma_{1}t}
\end{equation}

\begin{figure}[h]
\includegraphics[width=7cm]{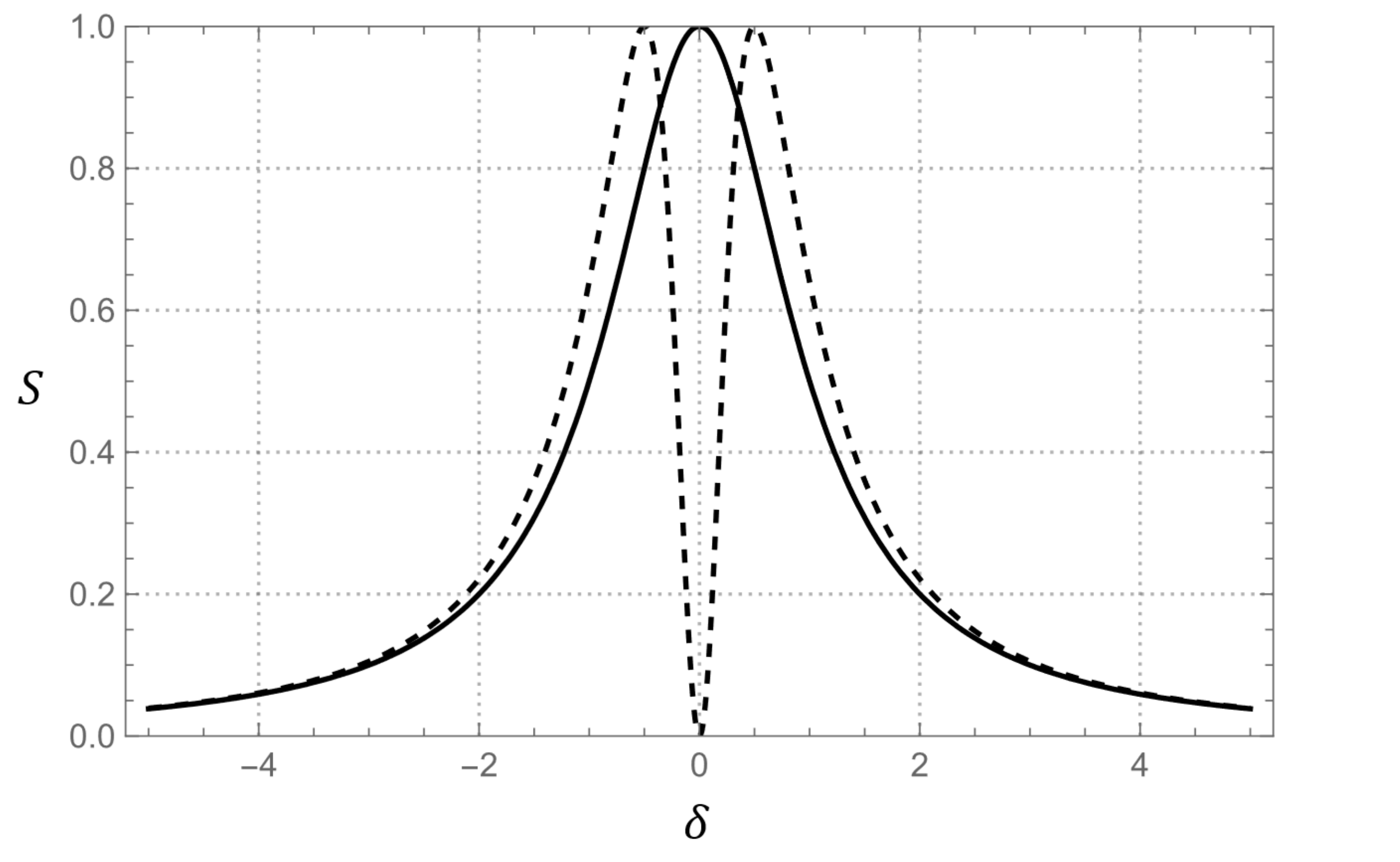}
\caption{The modified line shape (24) (dashed) as a function of $\delta/\gamma_{1}$ and its comparison with
the Breit-Wigner line shape (solid).}
\centering
\end{figure}

It is again interesting to note that the time scale is governed by $\gamma_1/2$ rather than $\gamma_1$.

The basic idea presented above is easily extended to more complex situations.
For example, two-photon decay in the system as shown in figure 4 which is easily realizable
atoms and molecules. The full Hamiltonian for this system can be written as
\begin{align}
H&=\hbar\omega_{13}\ket{1}\bra{1}+\hbar\omega_{23}\ket{2}\bra{2}+\hbar\omega_{43}\ket{4}\bra{4} \notag \\
&-\hbar(G_{l}e^{-i\omega_{l}t}\ket{4}\bra{2}+h.c.) \notag \\
&+\sum_{ks}\hbar\omega_{ks}a_{kx}^{\dagger}a_{ks}+V_{12}+V_{23}+V_{42}
\end{align}
where the meaning of different terms is obvious. Again a canonical transformation will
change the above $H$ into a time-independent $H$.
For $V_{42}\rightarrow0$, the above Hamiltonian becomes identical to the one for the quantum field theoretic extended cascade field model.
We thus have a simple atomic realization of the field-theoretic model. As shown recently
[7], this system exhibits very interesting two photon absorption characteristics. Clearly, the
electromagnetic coupling between the levels $\ket{2}$ and $\ket{4}$ can produce a double pole in the
decay of the system. It is interesting that a system equivalent to this has been studied by
Bhamathi and Sudarshan [2].

\begin{figure}[h]
\includegraphics[width=7cm]{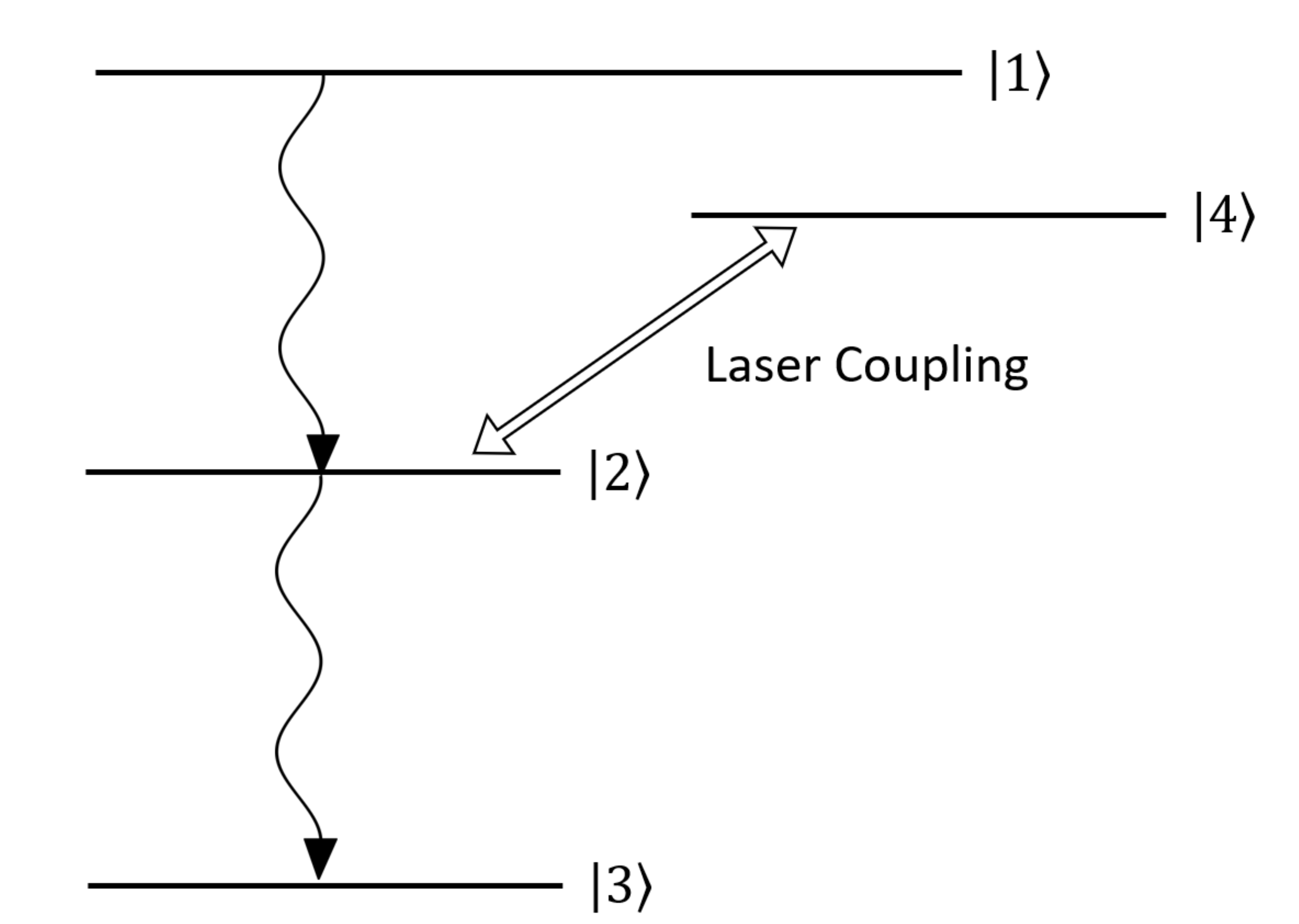}
\caption{A scheme involving laser coupling the intermediate state $\ket{2}$ which will create pole of
order two in the two-photon decay. This provides an analog of the extended cascade model.}
\centering
\end{figure}

\section{DOUBLE POLES AND INTERFERENCE EFFECTS}

The existence of double poles and the possibility of a line shape which is a derivative of Lorentzian suggest that the quantum interferences must be crucial.
This is indeed the case as can be seen from the following considerations.
The electromagnetic coupling of $\ket{1}$ and $\ket{2}$ produced dressed states $\ket{\psi_{\pm}}=\frac{1}{\sqrt{2}}(\pm\ket{1}+\ket{2})$ with eigenvalues
$\pm G_l$. Since $G_{l}\sim\gamma$, the two states are within the radiative line width. We pump the
population in the state $\ket{1}$ which is equivalent to pumping in both $\ket{\psi_{\pm}}$ as $\ket{1}=(\ket{\psi_{+}}+\ket{\psi_{-}})/\sqrt{2}$. Both states $\ket{\psi_{\pm}}$ can decay to $\ket{3}$ as $\ket{\psi_{\pm}}$ involve admixtures of $\ket{1}$ and $\ket{2}$. 
These two decays will not be independent [8,9] as $\overrightarrow{d}_{+3}\cdot\overrightarrow{d}_{-3}^{*}\neq0$ and as $G_{l}\sim\gamma$.

\section{EXPONENTIAL DECAY RECOVERED}

We also examine the initial conditions for our system which would result in exponential
decay. From Eq. (16) it is seen that
\begin{align}
&\frac{d}{d\tau}\left\langle (A_{13}(t+\tau)+iA_{23}(t+\tau))A_{31}(t)\right\rangle  \notag \\
&+\frac{\gamma_{1}}{2}\left\langle (A_{13}(t+\tau)+iA_{23}(t+\tau))A_{31}(t)\right\rangle =0
\end{align}
if $G_{l}=\frac{\gamma_{1}}{2},\gamma_{2}\rightarrow0$. Thus the correlation function defined in terms of the vector
$\tilde{\psi}=\frac{1}{\sqrt{2}}(\ket{1}+i\ket{2})$ obeys simple exponential decay law with a time scale governed by $\gamma_{1}/2$ rather than $\gamma_{1}$:
\begin{equation}
\left\langle A_{\tilde{\psi}3}(t+\tau)A_{3\tilde{\psi}}(t)\right\rangle =e^{-\gamma_{1}\tau/2}\left\langle A_{\tilde{\psi}\tilde{\psi}}(t)\right\rangle 
\end{equation}
Thus a pumping of the system to the state $\tilde{\psi}$ rather than $\ket{1}$ will result in exponential decay[10].

Thus, in conclusion, we have shown how higher-order poles in the decay of states can
be produced by using resonant electromagnetic fields. We demonstrated this by creating
a pole of order two. Clearly, the technique is quite versatile and by using combinations of
electromagnetic fields we can create poles of higher order.

I thank George Sudarshan for discussions on higher order poles of S-Matrix and R.P.
Singh for help in preparation of this paper.

\end{document}